\documentclass[showpacs,prl,amsmath,amssymb,superscriptaddress,nobibnotes,twocolumn, preprintnumbers]{revtex4-1}
\usepackage{graphicx}
\usepackage{endnotes}
\usepackage{bm}
\usepackage{epsfig}
\usepackage{amsmath}
\usepackage{color}

\newcommand{\eg}{{\it e.g.}}

\newcommand{\BiS}{LaO$_{0.5}$F$_{0.5}$BiS$_2$}

\newcommand{\parallelsum}{\mathbin{\!/\mkern-5mu/\!}}

\begin{document}
%


\title{Anisotropic two-gap superconductivity and the absence of a Pauli paramagnetic limit in single-crystalline LaO$_{0.5}$F$_{0.5}$BiS$_2$}

\author{Y.~C.~Chan$^\ddagger$}
\author{K.~Y.~Yip$^\ddagger$}
\author{Y.~W.~Cheung}
\author{Y.~T.~Chan}
\author{Q.~Niu}
\affiliation{Department of Physics, The Chinese University of Hong Kong, Shatin, New Territories, Hong Kong, China}

\author{J.~Kajitani} 
\author{R.~Higashinaka} 
\author{T.~D.~Matsuda}
\affiliation{Department of Physics, Tokyo Metropolitan University, Hachioji, Tokyo 192-0397, Japan}

\author{Y.~Yanase}
\affiliation{Department of Physics, Kyoto University, Kyoto 606-8502, Japan}

\author{Y.~Aoki}
\affiliation{Department of Physics, Tokyo Metropolitan University, Hachioji, Tokyo 192-0397, Japan}

\author{K.~T.~Lai}
\affiliation{Department of Physics, The Chinese University of Hong Kong, Shatin, New Territories, Hong Kong, China}
\author{Swee~K.~Goh}
\email{skgoh@phy.cuhk.edu.hk}
\affiliation{Department of Physics, The Chinese University of Hong Kong, Shatin, New Territories, Hong Kong, China}
\affiliation{Shenzhen Research Institute, The Chinese University of Hong Kong, Shatin, New Territories, Hong Kong, China}
\date{\today}


\begin{abstract}
Ambient-pressure-grown LaO$_{0.5}$F$_{0.5}$BiS$_2$ with a superconducting transition temperature $T_{c}~\sim$~3~K possesses a highly anisotropic normal state. By a series of electrical resistivity measurements with a magnetic field direction varying between the crystalline $c$-axis and the $ab$-plane, we present the first datasets displaying the temperature dependence of the out-of-plane upper critical field $H_{c2}^{\perp}(T)$, the in-plane upper critical field $H_{c2}^{\parallelsum}(T)$, as well as the angular dependence of $H_{c2}$ at fixed temperatures for ambient-pressure-grown LaO$_{0.5}$F$_{0.5}$BiS$_2$ single crystals. The anisotropy of the superconductivity, $H_{c2}^{\parallelsum}/H_{c2}^{\perp}$, reaches $\sim$16 on approaching 0~K, but it decreases significantly near $T_{c}$. A pronounced upward curvature of $H_{c2}^{\parallelsum}(T)$ is observed near $T_{c}$, which we analyze using a two-gap model. Moreover, $H_{c2}^{\parallelsum}(0)$ is found to exceed the Pauli paramagnetic limit, which can be understood by considering the strong spin-orbit coupling associated with Bi as well as the breaking of the local inversion symmetry at the electronically active BiS$_2$ bilayers. Hence, LaO$_{0.5}$F$_{0.5}$BiS$_2$ with a centrosymmetric lattice structure is a unique platform to explore the physics associated with local parity violation in the bulk crystal.
\end{abstract}


\maketitle

\section{Introduction}

The recent discovery of superconductivity in compounds containing BiS$_2$ layers \cite{Mizuguchi2012a, Singh2012, Yazici2015} has quickly inspired comparisons with other well-known layered superconductors, such as cuprates \cite{bednorz1986, wu1987} and Fe-based systems \cite{Kamihara2008, Rotter_2008}. The structural similarity stems from the fact that the BiS$_2$ layers are separated from each other by some block layers, which can be chemically manipulated to induce superconductivity. For instance, the insulating parent compound $R$OBiS$_2$ ($R$ is a rare-earth element) can be made superconducting through the partial substitution of O by F \cite{Yazici2013a, Mizuguchi2014a, Morice2016}. This partial substitution introduces extra electrons onto the BiS$_2$ layers, making the system more metallic before the realization of a superconducting ground state at low temperatures. Additionally, the superconducting state can be induced via a partial substitution of $R$ with tetravalent elements such as Th, Hf, Zr or Ti \cite{Yazici2013}. Given the flexibility of chemically manipulating the block layers to induce superconductivity, more BiS$_2$-based superconductors with a variety of block layers can be expected.

LaO$_{1-x}$F$_{x}$BiS$_2$ is one of the most heavily studied BiS$_2$-based series \cite{Mizuguchi2012, Kotegawa2012, Deguchi2013, Lee2013, Higashinaka2014, Nagao2015}. With an increasing F concentration $x$, superconductivity appears at $x\geq0.2$, and the superconducting transition temperature ($T_c$) reaches a maximum of $\sim$3~K at $x=0.5$ \cite{Mizuguchi2014a, Mizuguchi2012}. With the application of an external pressure of around 1~GPa, $T_c$ of \BiS\ can be rapidly enhanced to $\sim$10~K \cite{Kotegawa2012, Jha2015, Tomita2014}. Interestingly, \BiS\ synthesized under high pressure can superconduct at 10.5~K even at ambient pressure \cite{Kotegawa2012, Mizuguchi2012, Mizuguchi2014a}, which represents the highest $T_c$ among all BiS$_2$-based superconductors discovered thus far. To distinguish between two variants of \BiS, we denote the ambient-pressure-grown and high-pressure-annealed samples as AP-\BiS\ and HP-\BiS, respectively.

The layered nature of BiS$_2$-based systems naturally raises the question concerning the anisotropy of the electronic and superconducting properties. Band structure calculations show that the Fermi surface is cylindrical with a negligible $k_z$ dependence and a strong nesting at $(\pi, \pi, 0)$ \cite{Wan2013, Yildirim2013, Usui2012}, indicating a highly anisotropic electronic structure. To extract the anisotropy of the superconductivity, upper critical fields ($H_{c2}$) under varying temperatures and applied magnetic field directions are powerful probes. For HP-\BiS, the upper critical field anisotropy was inferred by analyzing the temperature derivative of the electrical resistivity of polycrystals, resulting in an anisotropy factor $\gamma$ of 7.4 \cite{Mizuguchi2014}, where $\gamma=H_{c2}^{\parallelsum}/H_{c2}^{\perp}$ with $H_{c2}^{\parallelsum}$ ($H_{c2}^{\perp}$) being the in-plane (out-of-plane) upper critical field. For AP-\BiS, while single crystals have been available for quite some time, the construction of the temperature-field phase diagrams with different field directions is surprisingly absent. 

In this article, we present the first datasets showing $H_{c2}$ of AP-\BiS\ measured at different field directions, constructed by measuring the electrical resistivity of single crystals. Our data indicate that the superconductivity is highly anisotropic. Furthermore, $H_{c2}^{\parallelsum}(T)$ exhibits a pronounced upward curvature near $T_c$, and the $H_{c2}^{\parallelsum}(0)$ is enhanced well beyond the Pauli paramagnetic limit $H_{p}$.

\section{Experimental}
\begin{figure}[!t]\centering
      \resizebox{8.5cm}{!}{
              \includegraphics{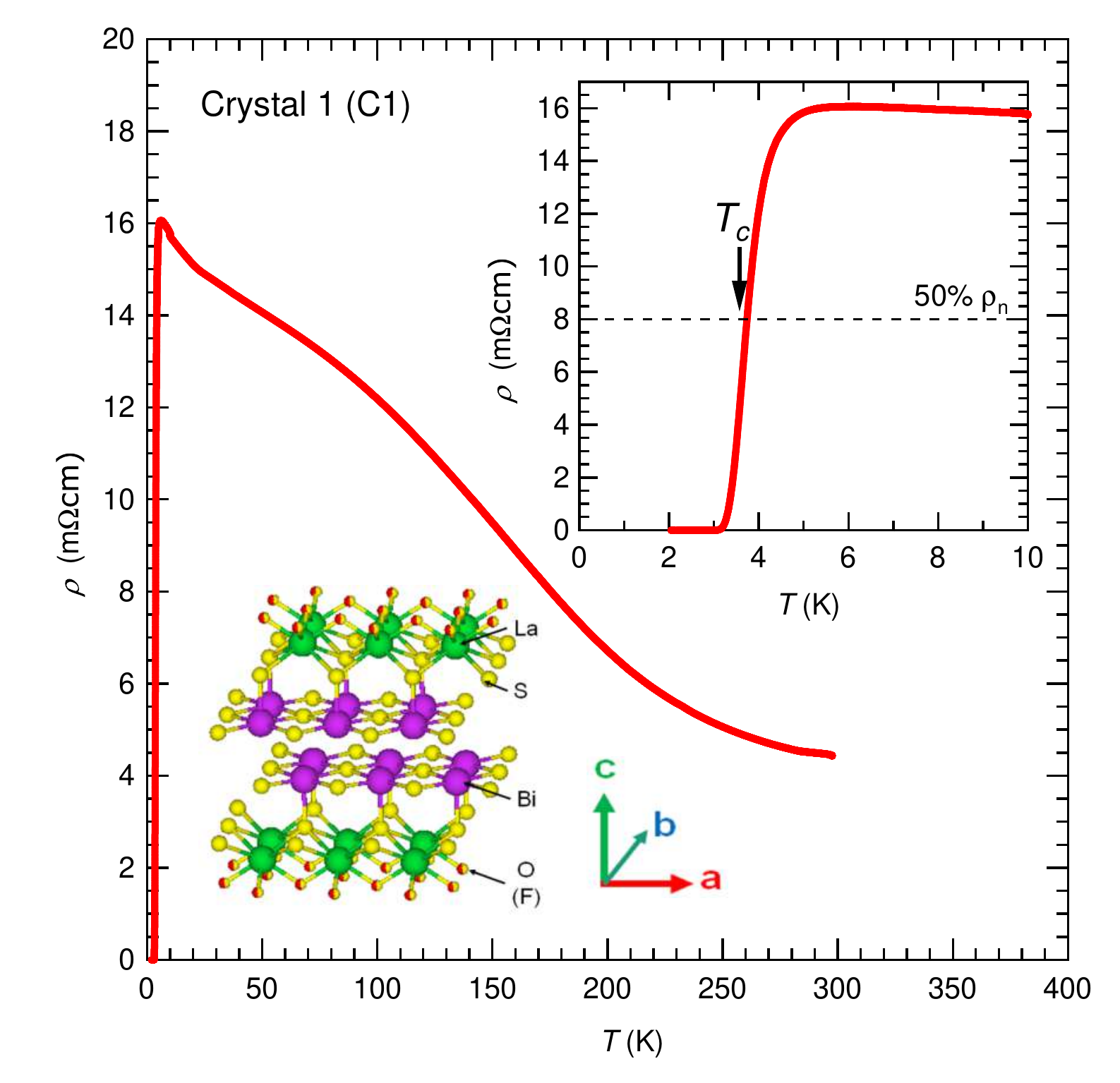}}       				
              \caption{\label{fig1} (Color online) Temperature dependence of resistivity $\rho(T)$ at ambient pressure without a magnetic field for Crystal~1.  Top inset: the enlarged resistivity curve near the superconducting transition, with the arrow indicating the ${T_{c}}$, defined as 50\% of the normal state resistivity $\rho_n$. Bottom inset: The crystal structure of LaO$_{0.5}$F$_{0.5}$BiS$_2$ with a space group of $P4/nmm$. }
\end{figure}

Single crystals of LaO$_{0.5}$F$_{0.5}$BiS$_2$ were grown by the CsCl flux method using stoichiometrically-mixed starting materials consisting of La$_2$S$_3$ (99.9~\%), Bi$_2$O$_3$ (99.999~\%), BiF$_3$ (99.99~\%) powders, as well as Bi (99.99~\%) and Bi$_2$S$_3$ (99.999~\%) grains. They were homogenized with the flux medium CsCl (99.9~\%) and were then sealed in a quartz tube under a vacuum of $1\times10^{-3}$~Pa. The ampoule was annealed at 900$^\circ$C for 12 hours, followed by a slow cooling to 500$^\circ$C at a rate of 2.4$^\circ$C/h. After the heat treatment, the flux was removed by H$_2$O to extract single crystals. 
Single-crystal and powder X-ray analyses have shown that the crystal structure belongs to the space group $P4/nmm$ (c.f. Fig.~\ref{fig1}) and the lattice parameters are $a=4.0585$~\AA\ and $c=13.324$~\AA, which is consistent with previous reports \cite{Mizuguchi2012}. Assuming Vegard's law \cite{Nagao2015}, the value of $c$ indicates $x=0.502\pm0.029$.
The temperature dependence of electrical resistivity $\rho (T)$ was measured using a standard four-probe technique with current flowing in the $ab$-plane. The electrical contacts were made with gold wires glued on a freshly cleaved surface of the sample by silver paste (Dupont 6838). Crystal~1 (C1) was measured down to 30~mK using a dilution fridge (BlueFors Cryogenics) equipped with a 14~T magnet, whereas Crystal~2 (C2) was studied using a rotator in a Physical Property Measurement System (Quantum Design) down to 2.0~K.

\section{Results and Discussion}

Fig. \ref{fig1} shows $\rho(T)$ for C1 from 300~K to 2~K at ambient pressure without a magnetic field. The $\rho(T)$ curve exhibits a semiconducting-like behavior, which broadly resembles that of polycrystalline samples \cite{Tomita2014, Kotegawa2012, Jha2015, Wolowiec2013a}, except a convex curvature from $ \sim $ 15 K to 200 K. The convex curvature could be related to the discovery of weak superlattice reflections at low temperatures by single-crystal X-ray diffraction \cite{private}. The top inset shows the enlarged $\rho(T)$ curve from 2 K to 10 K, featuring a sharp superconducting transition with the onset temperature $ \sim $ 5 K. For the quantitative analysis of the superconductivity, we adopt the `50\% criterion', as indicated by the arrow in top inset of Fig.~\ref{fig1}, by defining $T_c$ ($H_{c2}$) as the temperature (field) at which the resistivity is 50\% of the normal state value $\rho_n$. With this criterion, $T_c$ of C1 is 3.7~K.

\begin{figure}[!t]\centering
       \resizebox{8.5cm}{!}{
              \includegraphics{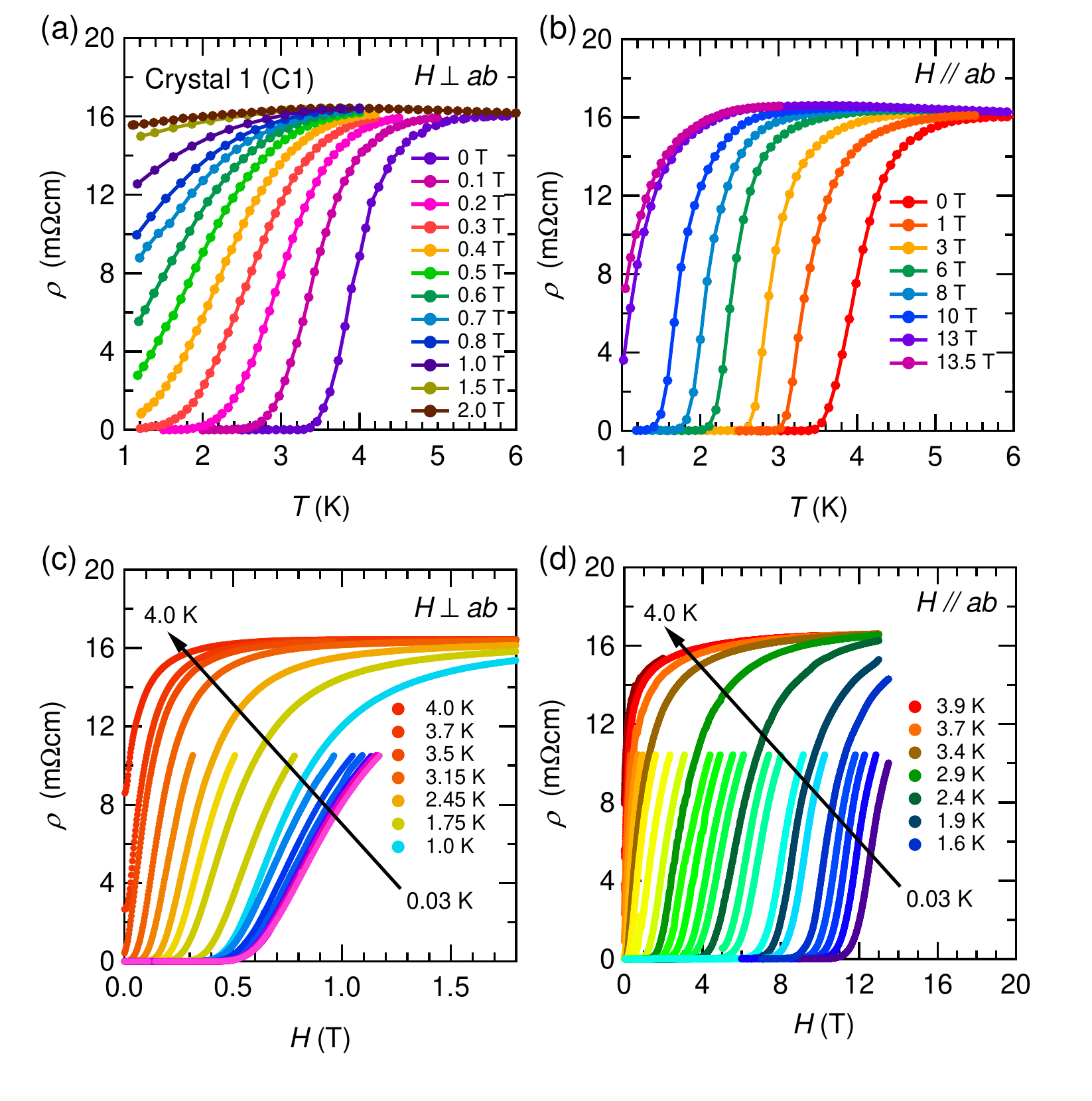}}                				
              \caption{\label{fig2} (Color online) Temperature dependence of resistivity $\rho(T) $ for Crystal~1 under various magnetic fields with (a) $H\perp ab$ and (b) $H\parallelsum ab$. Field dependence of resistivity $\rho(H) $ at different temperatures with (c) $H\perp ab$ and (d) $H\parallelsum ab$. The black arrow indicates increasing temperatures. Representative curves covering a longer field range are labelled in the legend. 
              }
\end{figure}

 Figs. \ref{fig2} (a) and (b) show $\rho(T) $ for C1 under various magnetic fields with $H\perp ab$ and $H\parallelsum ab$, respectively. For $H\perp ab$, $T_c$ is greatly suppressed with an increasing magnetic field. On the contrary, for $H\parallelsum ab$, $T_c$ is significantly more robust against the applied  field, indicating a large anisotropy factor $\gamma$ for \BiS. At low temperatures, the superconducting transition becomes significantly broader for $H\perp ab$. The broadening can be quantified by $\Delta T_c/T_c = [ T_c(90\%\rho_n) -T_c (10\%\rho_n)] /T_c (50\%\rho_n)$. For $H\parallelsum ab$, $\Delta T_c/T_c$ merely increases from $\sim$0.26 at $T_c$ to 0.46 at $0.44T_c$. However, for $H\perp ab$, $\Delta T_c/T_c$ already reaches 0.82 at $0.56T_c$. These in-field behaviours could be due to the poor pinning of pancake vortices in this anisotropic, two-dimensional superconductor. This scenario is plausible, since $\xi_{\perp}$ (see below) is less than the lattice constant $c$. 
 
 Figs.~\ref{fig2} (c) and (d) display the field dependence of electrical resistivity $\rho(H) $ at different temperatures from 4.0~K down to $\sim$ 30 mK with $H\perp ab$ and $H\parallelsum ab$, respectively. With increasing $H$ for $H\perp ab$, $\rho(H)$ increases and, at 4.0~K and 1.0~K, saturates at the same field independent normal state value of 16.0~m$\Omega$cm. This shows that the magnetoresistance does not vary strongly within the field and temperature windows of interest. With $H\parallelsum ab$, the magnetoresistance is expected to be even weaker. In fact, the weak magnetoresistance is not surprising, given the large residual resistivity of the sample. Hence, we take 16.0~m$\Omega$cm as the universal normal state resistivity for the determination of $H_{c2}$.
 Again, $H_{c2}$ determined from $\rho(H)$ is sensitive to the field direction (note the range of field axes in Figs.~\ref{fig2} (c) and (d)).

\begin{figure}[!t]\centering
       \resizebox{8.5cm}{!}{
              \includegraphics{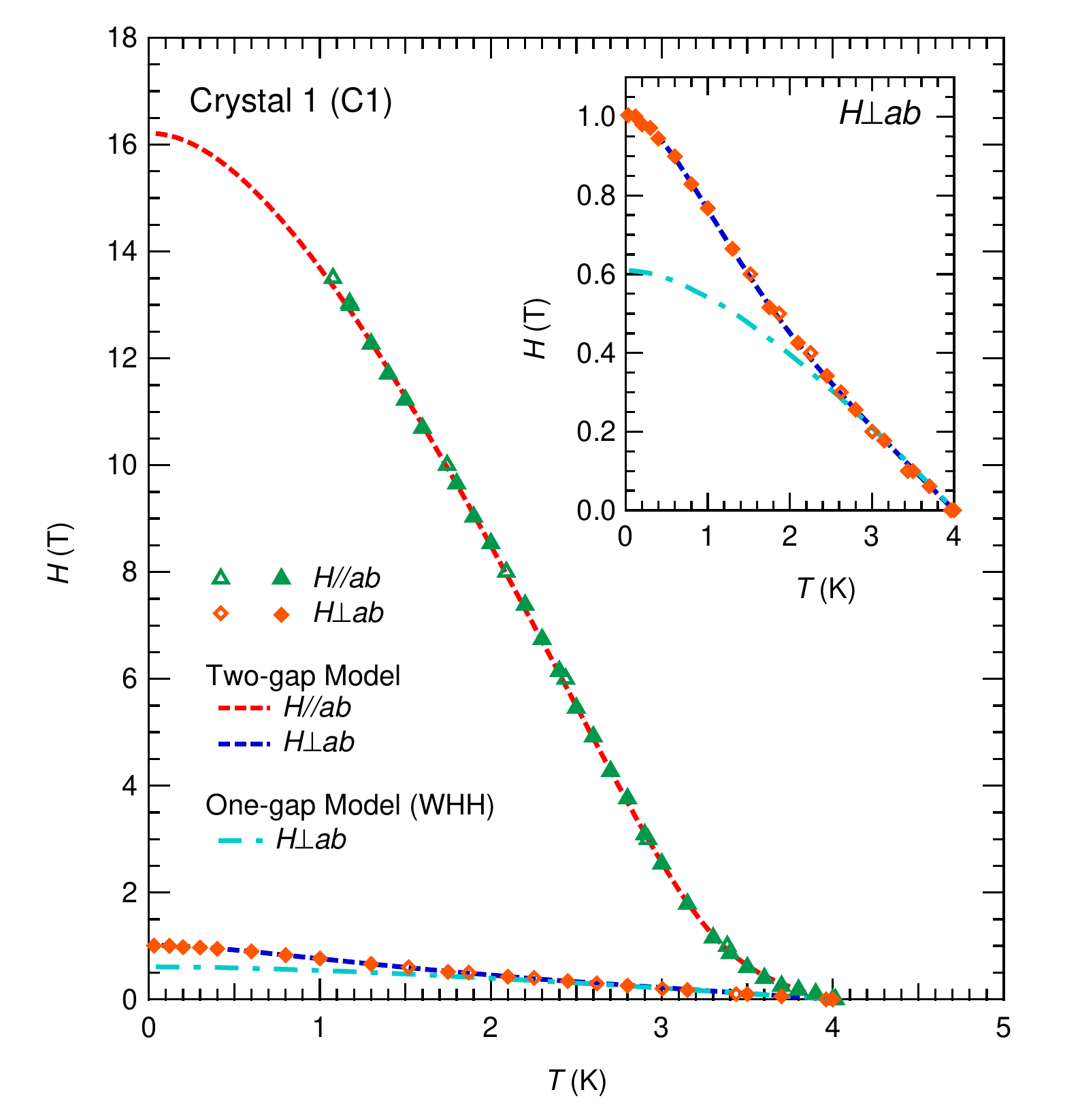}}                				
                    \caption{\label{fig3} (Color online)  Temperature dependence of $H_{c2}$ with $H\parallelsum ab$ and $H\perp ab$ for Crystal~1. Experimental data, shown in symbols, are fitted with the two-gap model (dashed line) and the Werthamer-Helfand-Hohenberg (WHH) theory (dash-dotted line). Inset: The enlargement of the low-field region for a clearer view of $H_{c2}^{\perp}(T)$.}
\end{figure}

From $\rho(T)$ and $\rho(H)$, we construct the $H-T$ phase diagram for $H\parallelsum ab$ and $H\perp ab$, as displayed in Fig.~\ref{fig3}. The critical values obtained from both temperature sweeps (open symbols) and field sweeps (closed symbols) overlap smoothly, exhibiting an excellent agreement with each other. From these data, $\xi_{\parallelsum}$ and $\xi_{\perp}$ can be calculated by using $H_{c2}^{\perp}=\dfrac{\phi_0}{2\pi \xi_{\parallelsum}^2}$ and $H_{c2}^{\parallelsum}=\dfrac{\phi_0}{2\pi \xi_{\parallelsum} \xi_{\perp}}$, resulting in  $\xi_{\parallelsum} \sim$ 21.7~nm and $\xi_{\perp} \sim$ 1.13~nm at 1 K.  
Our single crystal data further reveal two interesting behaviors that have not been reported in polycrystalline LaO$_{0.5}$F$_{0.5}$BiS$_2$ studies \cite{Jha2015, Higashinaka2014,  fang2016}. First, $H_{c2}^{\parallelsum}(T)$ becomes unexpectedly high when approaching 0~K. 
Below 2~K, it exceeds the Pauli limited field $H_{p}(0)$~[T]~=~1.84~$T_{c}$~[K] $\approx$ 7.36~T for weakly coupled BCS superconductors \cite{Clogston1962, Chandrasekhar1962}. Second, a pronounced upward curvature of $H_{c2}^{\parallelsum}(T)$ near $T_{c}$ is observed. These features are unambiguously not compatible with the one-gap Werthamer-Helfand-Hohenberg (WHH) theory \cite{Werthamer1966}. On the other hand, $H_{c2}^{\perp}(T)$ is much smaller and shows an upward curvature near 2 K (see the inset of Fig. \ref{fig3}). 
We have simulated the temperature dependence of $H_{c2}$ using the WHH theory (dash-dotted line), and the mismatch between the data and the simulation clearly implies that the WHH theory is also not applicable to $H_{c2}^{\perp}(T)$.

Muon spin relaxation measurements on AP-\BiS\ suggest the possibility of the two-gap superconductivity \cite{Zhang2016}. In addition, scanning tunneling spectroscopy data on NdO$_{0.5}$F$_{0.5}$BiS$_2$ reveal the existence of two superconducting gaps  \cite{Liu2014}. It has been pointed out that $H_{c2}(T)$ can be rather unconventional in the presence of multiple superconducting gaps, as discussed in MgB$_{2}$ and several iron-based superconductors \cite{Xing2017, Haenisch2015,Hunte2008}.
Inspired by these works, we apply the two-gap model \cite{Gurevich2003} to analyze the temperature dependence of $H_{c2}$ for LaO$_{0.5}$F$_{0.5}$BiS$_2$. Within the two-gap model in the dirty limit, $H_{c2}(T)$ is implicitly described by:
\begin{eqnarray}
a_{0}\left[\text{ln}t+U(h)\right]\left[\text{ln}t+U(\eta h)\right]  \nonumber \\
+a_{2}\left[\text{ln}t+U(\eta h)\right]
+a_{1}\left[\text{ln}t+U(h)\right]=0
\end{eqnarray}
with $a_{1}=1+(\lambda_{11}-\lambda_{22})/[(\lambda_{11}-\lambda_{22})^{2}+4\lambda_{12} \lambda_{21}]^{1/2}$, $a_{2}=1-(\lambda_{11}-\lambda_{22})/[(\lambda_{11}-\lambda_{22})^{2}+4\lambda_{12} \lambda_{21}]^{1/2}$,  
$a_{0}=2(\lambda_{11} \lambda_{22}-\lambda_{12} \lambda_{21})/[(\lambda_{11}-\lambda_{22})^{2}+4\lambda_{12} \lambda_{21}]^{1/2}$, $t=T/T_c$, $h=D_1H_{c2}/2\phi_{0} T$, $\eta=D_{2}/D_{1}$, and $U(x)=\psi(x+0.5)-\psi(0.5)$. $\lambda_{11} $ and  $\lambda_{22} $ are the intraband superconducting coupling constants, while $\lambda_{12} $ and $\lambda_{21} $ describe the interband coupling. $\phi_{0}$ is the magnetic flux quantum. $D_{1}$ ($D_{2}$) is the diffusion coefficient of Band 1 (Band 2) (Band 1 is defined as the band with the stronger coupling constant). $\psi(x)$ is the digamma function. The results of the fit are illustrated as the dashed lines in Fig.~\ref{fig3}, which clearly describe the experimental data nicely. Based on this model, $H_{c2}^{\parallelsum}(0)$ is estimated to be 16.2~T. Hence, $\gamma$ is as high as 16 as $T\rightarrow0$.

The diffusion coefficient $D_{2}$ extracted from the analysis is larger than $D_{1}$ for both  $H\perp ab$ and $H\parallelsum ab$. This explains the upward curvature of $H_{c2}$ for both field orientations.  
The smaller value for $D_{1}$ implies that Band 1 is dirtier \cite{Gurevich2007}, which could be the main contributing factor to the large enhancement of $H_{c2}$ at low temperatures. Indeed, previous studies \cite{orlando1979, Tarantini2011} have already demonstrated a significant enhancement of $H_{c2}$ by adding impurities and/or introducing defects to the superconducting systems. 

Whether or not our sample is in the dirty limit can be evaluated by comparing $\xi_{\parallelsum}$ with the Pippard coherence length $\xi_0$. Assuming a single cylindrical Fermi surface, the Fermi wavevector can be written as $k_F=\sqrt{2\pi nc}\simeq1.02\times10^9$~m$^{-1}$, where $c$ is the lattice constant and $n$ $\simeq$ $1.24\times 10^{20}$ cm$^{-3}$ at 10 K \cite{Awana2013}. The normal state resistivity of the sample does not vary much below 10~K (see top inset of Fig.~\ref{fig1}), therefore $n$ is not expected to change drastically at this temperature range. Take an effective mass $m^\star/m_e\approx 0.25$, estimated from Refs.~\cite{Wu2017, Sugimoto2015}, and $T_c=4$~K, we calculate that $\xi_0=0.18\times\frac{\hbar v_F}{k_BT_c}=0.18\times\frac{\hbar^2 k_F}{m^\star k_BT_c}\simeq163$~nm. 
Thus, $\xi_{\parallelsum}$ is much smaller than $\xi_0$. The mean free path $\ell$ can be estimated from $\xi_{\parallelsum}^{-1}=\xi_0^{-1}+\ell^{-1}$, leading to $\ell\simeq25$~nm. Hence, the criterion $\ell/\xi_0\ll1$ is satisfied even for a crude estimation \cite{mfpnote}, leading to the conclusion that the system is in the dirty limit. 

\begin{figure}[!t]\centering
       \resizebox{8.5cm}{!}{
              \includegraphics{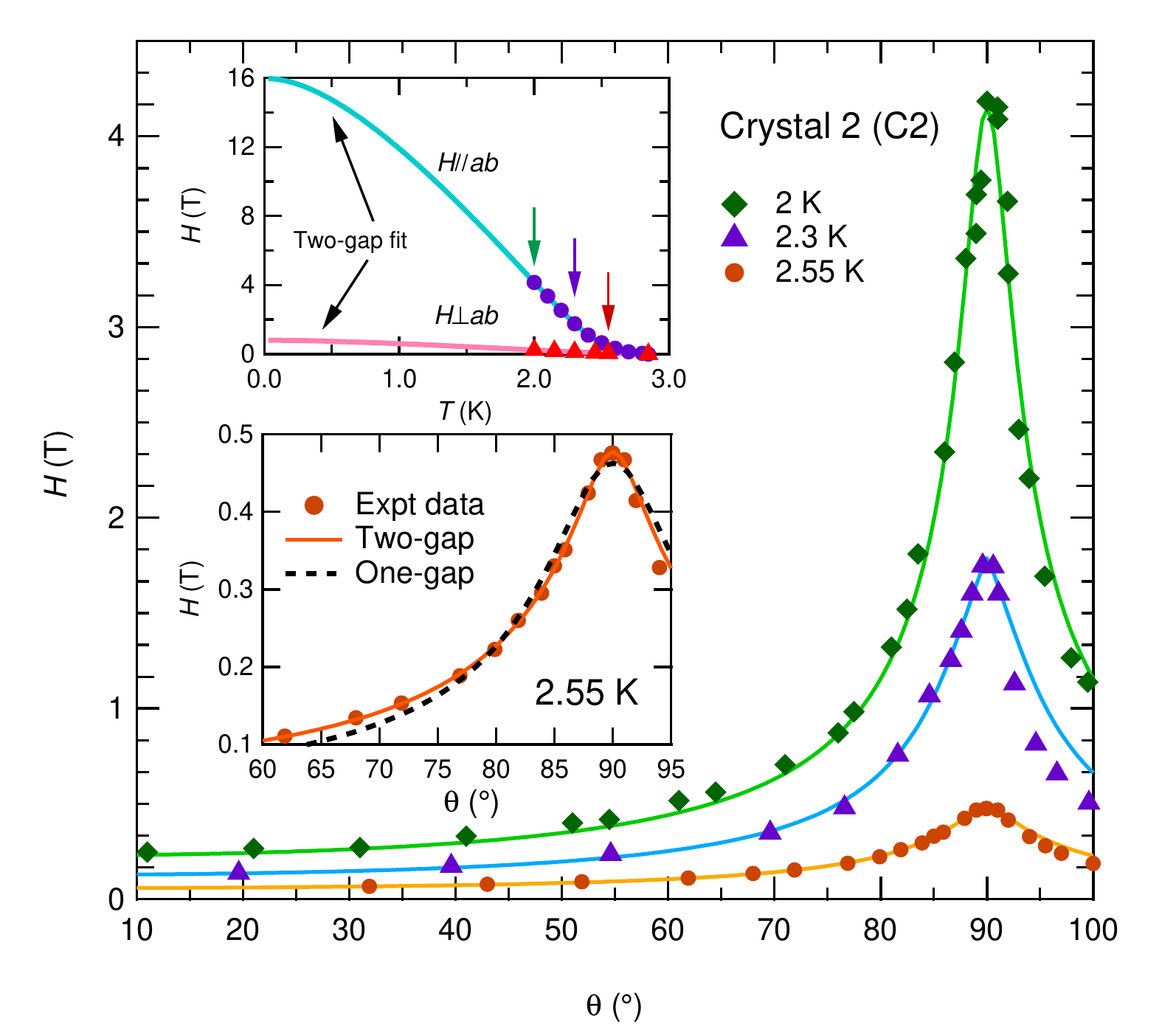}}                				
              \caption{\label{fig4} (Color online) Angular dependence of $H_{c2}$ for Crystal~2 at 2 K, 2.3 K and 2.55 K (solid symbols). The solid lines are the fits using the two-gap model. The angle $\theta=90^{\circ} (0^{\circ})$ corresponds to $H\parallelsum ab$ ($H\perp ab$). Top inset: $H-T$ phase diagram constructed for Crystal~2, showing a temperature dependence similar to the sample discussed earlier. Arrows indicate the temperatures where angular dependent studies were conducted. Bottom inset: Expanded low-field region for a clearer view of $H_{c2}(\theta)$ at 2.55~K.}
\end{figure}
Although the two-gap model in the dirty limit successfully captures the temperature dependence of  both $H_{c2}^{\parallelsum}$ and $H_{c2}^{\perp}$, the absence of the Pauli limit should be noted:  $H_{c2}^{\parallelsum}(0)$ exceed the Pauli field by a factor of two. Because of the large atomic number of Bi, the system has a strong spin-orbit coupling (SOC). A possible mechanism is spin-orbit scattering, in which the electron scattering depends on both its spin and orbital angular momentum in the presence of strong SOC. Theoretical calculations showed that such a spin-orbit scattering can enhance $H_{c2}^{\parallelsum}(0)$ by up to 5 times of $H_{p}$ \cite{Klemm1975}. Another mechanism is related to the spatial symmetry of the crystal structure. \BiS\ crystallizes in a centrosymmetric space group $P4/nmm$, which possesses a global inversion symmetry. However, upon examining the crystal structure of \BiS\ (Fig.~\ref{fig1}), the electronically active BiS$_2$ bilayers do not possess an inversion symmetry. Such a breaking of the local inversion symmetry can result in a large Rashba-Dresselhaus SOC \cite{Fischer2011, Maruyama2012, Caviglia2010, Goh2012, Shimozawa2014, Zhang2014, Wu2017}, which 
locks the spin directions onto the $ab$-plane. Hence the coupling between the external in-plane field and these spins is reduced, effectively protecting the Cooper pairs from depairing and consequently the Zeeman effect is suppressed.
Therefore, the combined effects from the dirty two-gap case and the strong spin-orbit coupling can explain the huge enhancement of $H_{c2}^{\parallelsum}(0)$ in \BiS. In the isostructural compounds LaO$_{0.5}$F$_{0.5}$BiSe$_2$ and LaO$_{0.5}$F$_{0.5}$BiSSe, a similar enhancement of $H_{c2}^{\parallelsum}$ at low temperatures has also been reported \cite{Shao2014, kase2017, Terui2017}, which can be understood using the same framework developed for our case. The anisotropy factor $\gamma$ at the 0~K limit is 13.6 and 32.3 for LaO$_{0.5}$F$_{0.5}$BiSe$_2$ and LaO$_{0.5}$F$_{0.5}$BiSSe, respectively. Note that other contributions to $H_{c2}$ enhancement may be possible, such as strong electron-phonon coupling and localized charge-density waves predicted by band structure calculations \cite{Yildirim2013, Li2013a, Wan2013, Schossmann1989, Morice2017}. Further investigations will shed light on this issue.
 

To further investigate the anisotropic superconductivity in single crystalline \BiS, we measured the angular dependence of $H_{c2}$ in Crystal 2 (C2) at 2~K, 2.3~K and 2.55~K. Although C2 has a lower $T_c$ of 2.84 K, the $H-T$ phase diagram is similar to C1 discussed earlier, as shown in the top inset of Fig.~\ref{fig4}. In the main panel of Fig.~\ref{fig4}, the collected $H_{c2}(\theta)$ data of C2 are displayed. $H_{c2}$ is highly sensitive to the field angle, and $H_{c2}(\theta)$ can be reasonably well-described by the two-gap model \cite{Gurevich2003}:
\begin{equation}
H_{c2}(\theta)=\frac{8\phi_{o}(T_{c}-T)}{\pi^2[a_{1}D_{1}(\theta)+a_{2}D_{2}(\theta)]}
\end{equation}
with the angular dependent diffusivities $D_{1}(\theta)$ and $D_{2}(\theta)$:  
\begin{equation}
D_{m}(\theta)=[D_{m}^{(a)2}\cos^2\theta+D_{m}^{(a)}D_{m}^{(c)}\sin^2\theta]^{1/2}; m=1,2
\end{equation}
where $D_m^{(a)}$ and $D_m^{(c)}$ are the principal values of diffusivity tensor in the $ab$-plane and along the $c$-axis, respectively. For this analysis, we also obtain $D_{2} > D_{1}$ for all angles $\theta$, consistent with the analysis of $H_{c2}(T)$ exhibited in the top inset of Fig.~\ref{fig4} and in Fig. \ref{fig3}.
For comparison, we also perform an analysis with a one-gap model by setting $a_2=0$, $D_{2}=0$ and $a_{1}=2$. The one-gap model, which is equivalent to the anisotropic mass model, gives a slightly poorer description of the data, as evidenced in the dataset at 2.55~K (bottom inset of Fig.4). Therefore, both the angular dependence and the temperature dependence of $H_{c2}$ in \BiS\ are in good agreement with the two-gap model.

Recently, laser-based angle-resolved photoemission spectroscopy (ARPES) on NdO$_{0.71}$F$_{0.29}$BiS$_2$ reported the existence of gap nodes on the same Fermi sheet, giving rise to superconducting gaps of distinct sizes \cite{Ota2017}. These gaps could be responsible for the two-gap physics we discussed above. A recent calculation discusses the possibility of having two different gaps on a single Fermi surface sheet, and shows that the gaps are connected to the variation of orbital mixing along the Fermi surface sheet \cite{Griffith2017}. However, thermal conductivity \cite{Yamashita2016} and penetration depth \cite{Jiao2015} measurements on the same Nd-based systems did not detect the existence of the gap nodes. It has been argued that the laser ARPES study probes a much smaller, and hence more homogeneous, region than the other probes. Similar studies on \BiS, in which we observe strong evidence of two-gap superconductivity, are highly desirable.

\section{Conclusions}
In summary, we have measured both the angular- and temperature-dependent $H_{c2}$ of LaO$_{0.5}$F$_{0.5}$BiS$_2$ single crystals. The temperature dependence of $H_{c2}$ at $H\parallelsum ab$ shows a pronounced upward curvature and at low temperatures, $H_{c2}^{\parallelsum}$ greatly exceeds the Pauli paramagnetic limit. We employ a dirty-limit two-gap model to describe our data, and discuss the role of spin-orbit coupling resulting from the breaking of the local inversion symmetry. The angular dependence of $H_{c2}$ can also be satisfactorily described with the two-gap model. Our data show that LaO$_{0.5}$F$_{0.5}$BiS$_2$ is a highly anisotropic two-gap spin-orbit coupled superconductor.

\section{Acknowledgments}
\begin{acknowledgments} We acknowledge Corentin Morice for discussion. This work was supported by Research Grants Council of Hong Kong (GRF/14301316, GRF/14300117), CUHK Direct Grant (No. 3132719, No. 3132720), CUHK Startup (No. 4930048), National Natural Science Foundation of China (No. 11504310), JSPS KAKENHI (JP15H03693, JP15K05178, JP16J05692, JP16K05454, JP15K05164 and JP15H05745), Grant-in-Aid for Scientific Research on Innovative Areas 
``J-Physics'' (JP15H05884) and 
``Topological Materials Science'' (JP16H00991)  \end{acknowledgments}

$^\ddagger$Y.C.C. and K.Y.Y. contributed equally to this work.


\begin{thebibliography}{55}
\bibitem{mfpnote} Direct estimation of the mean free path using a single band Drude model and the measured resistivity value lead to $\ell_\rho\simeq0.34$~nm. This value is much smaller than $\ell$ estimated from $\xi_0$ and $\xi_{\parallelsum}$. Such a small $\ell_\rho$, even slightly smaller than the lattice constant $a$, is puzzling. The small $\ell_\rho$ could come from an overestimation of the effective sample thickness, due to the fact that LaO$_{0.5}$F$_{0.5}$BiS$_2$ is a highly two-dimensional material, and hence the transport current only flows in the top few layers. Furthermore, the estimation assumes a single, strictly two-dimensional Fermi surface and the absence of any Fermi surface gapping (\eg\ due to charge order). A more realistic estimation could change the values of $\ell_\rho$, but unlikely to affect the conclusion that both $\ell$ and $\ell_\rho$ are much smaller than $\xi_0$.
\end{thebibliography}


 
%
%

\end{document}